\documentclass[12pt,english]{article}
\usepackage[T1]{fontenc}
\usepackage[latin9]{inputenc}
\usepackage{array}
\usepackage{float}
\usepackage{multirow}
\usepackage{amsmath}
\usepackage{amsthm}
\usepackage{amssymb}
\usepackage{graphicx}
\usepackage{geometry}
\usepackage{setspace}
\usepackage[authoryear]{natbib}
\usepackage{hyperref}

\makeatletter

\providecommand{\tabularnewline}{\\}

\theoremstyle{definition}
\newtheorem{defn}{\protect\definitionname}
\theoremstyle{plain}
\newtheorem{prop}{\protect\propositionname}

\usepackage{tikz}
\usepackage{pgfplots}
\usepackage{xfrac}
\usepgfplotslibrary{fillbetween}
\setcitestyle{aysep={}}

\makeatother

\usepackage{babel}
\providecommand{\definitionname}{Definition}
\providecommand{\propositionname}{Proposition}

\begin{document}
\title{A Flexible Measure of Voter Polarization\thanks{I thank Annal� Casanueva, Edoardo Cefala, Philipp Denter, Mikhail Drugov, Robert Somogyi, and audience at EPCS Riga for helpful comments. Financial support from MICIN/AEI/10.13039/501100011033 grants CEX2021-001181-M, PID2020-118022GB-I00, and RYC2021-032163-I; and Comunidad de Madrid grants EPUC3M11 (V PRICIT) and H2019/HUM-5891 is gratefully acknowledged.}}
\author{Boris Ginzburg\thanks{Department of Economics, Universidad Carlos III de Madrid, Calle Madrid 126, 28903 Getafe (Madrid), Spain. Email: bginzbur@eco.uc3m.es.}}
\maketitle
\begin{abstract}
This paper introduces a definition of ideological polarization of an electorate around a particular central point. By being flexible about the location or width of the center, this measure enables the researcher to analyze polarization around any point of interest. The paper then applies this approach to US voter survey data between 2004 and 2020, showing how polarization between right-of-center voters and the rest of the electorate was increasing gradually, while polarization between left-wingers and the rest was originally constant and then rose steeply. It also shows how, following elections, polarization around left-wing positions decreased while polarization around right-wing positions increased. Furthermore, the paper shows how this measure can be used to find cleavage points around which polarization changed the most. I then show how ideological polarization as defined here is related to other phenomena, such as affective polarization and increased salience of divisive issues.

Keywords: polarization, ideology.
\end{abstract}
\newpage

Mass polarization is often described as ``disappearance of the center'' \citep{Abramowitz2010center}, or ``movement from the center toward the extremes'' \citep{fiorina2008polarization,levendusky2011red}. Given the significant consequences that polarization entails for political systems, it is not surprising that it has become a focus of much research. Yet detecting such a shift in policy preferences is difficult, because there is no single way of comparing distributions of voters' political positions. For example, much of the recent debate over whether the US electorate is becoming more ideologically polarized (see \citealt{fiorina2008polarization,abramowitz2008polarization,westfall2015perceiving,menchaca2023americans}) is due to the different definitions of polarization \citep{lelkes2016mass,mehlhaff2022group}. 

When voters have clear group identities, such as party affiliations, polarization may be evaluated by measuring ideological distance between groups and the degree of homogeneity within them \citep{esteban1994measurement,duclos2004polarization,mehlhaff2022group}, or the degree of overlap between group-specific distributions of voters' positions \citep{levendusky2011red,lelkes2016mass}. However, such indicators are of little help when voters' group identities are not clearly defined, or when one needs to measure ideological polarization across a heterogeneous electorate, rather than between specific groups.

%\citep{montagnes2019bounding,egan2020identity,orr2023affective}

To measure mass polarization without relying on exogenous group identities, one needs a measure of dispersion around a certain central point. For this, many papers focus on statistical moments of the distribution of voters' positions. These include variance or standard deviation \citep{dimaggio1996have,ezrow2007variance,levendusky2009microfoundations}; kurtosis \citep{dimaggio1996have,baldassarri2007dynamics}; or bimodality coefficient, which is a function of skewness and kurtosis \citep{lelkes2016mass}. By construction, all these indicators measure polarization as dispersion around the mean. For some applications, it is reasonable to assume that the mean is the relevant central point. But for others, it is not. For example, in formal models that follow \citet{Downs_Anthony1957}, it often makes more sense to think of dispersion around the median. In models of information acquisition or manipulation (see \citealp{rosenfeld2024information,denter2023troll,ginzburg2025fact}), the central point might be the position of a voter who is undecided in the absence of further information. When working with survey data, a researcher may think that a centrist is a voter who gives a ``neutral'' response -- for example, who neither likes nor dislikes a certain policy proposal or a politician. In all of these cases, the centrist position is not, in general, the average position.

Other studies use survey data to measure polarization through a change in the share of voters that identify as centrists. \citet{fiorina2008political} and \citet{iversen2015information} consider as centrists those voters who placed themselves exactly in the middle of, respectively, a 7-point and an 11-point left-right scale. \citet{levendusky2009microfoundations} focuses on those who place themselves in a wider 3-5 interval on a 7-point scale. \citet{abramowitz2008polarization} label as moderates those with a score of 2 or 3 on a 0-7 aggregate scale. Thus, the definition of centrists can vary, and such measures require the researcher to define how far the left and the right boundaries of the centrist interval are. Consequently, the location of these boundaries can have a considerable effect on the results of the analysis.

In this paper, I propose a way of evaluating polarization that operationalizes the idea of the disappearance of the center. Crucially, the model neither relies on a specific central position nor makes assumptions about the width of the relevant interval around it. Instead, it allows the researcher to select the central position depending on the application. Thus, rather than seeing polarization as a single characteristic of the distribution of voters, the model measures polarization \emph{around any particular position} that a researcher finds relevant.

As a starting point, I take an electorate characterized by a distribution of political positions over a unidimensional policy space. A single point $x^{*}$ -- which may be any position relevant to the researcher -- is taken as the central position. I then introduce a partial order of distributions in terms of polarization around $x^{*}$. Under that order, polarization is higher if the share of voters belonging to \emph{any} interval that includes $x^{*}$ is smaller. Thus, polarization increases if the fraction of centrist voters falls, regardless of how restrictively one defines the centrists. The paper then proves a simple necessary and sufficient condition that can be easily applied to a pair of distributions to determine whether one represents greater polarization around a particular position $x^{*}$. It then introduces a numerical indicator, $P\left(F,x^{*}\right)$, that measures, in a manner consistent with the above ordering, polarization around $x^{*}$ of a given distribution $F$ of voters' positions. This numerical measure can be used to compare any two distributions in terms of polarization.

Using US survey data as an illustration, I then show how this measure can contribute to our understanding of polarization in two ways. First, by being flexible around the location of the central point $x^{*}$, the $P\left(F,x^{*}\right)$ index can be used to measure polarization around any point that a researcher finds relevant. This enables a researcher to gain more detailed insights about evolution of polarization, compared to what can be obtained with more traditional mean-centered indicators such as variance. In particular, the US data shows that between 2004 and 2020, polarization around various right-of-center positions was gradually growing. To put it differently, throughout this period right-wing voters -- defined variously depending on the dividing line between the right and the rest -- were becoming progressively more distant ideologically from the rest of the electorate. On the other hand, during the same period polarization around left-of-center positions was roughly unchanged until 2012, and increased rapidly afterwards. 

Second, I show how the measure $P\left(F,x^{*}\right)$ can be used to evaluate changes in polarization over shorter periods, for example, before and after an election. A researcher comparing pre- and post-election surveys and using variance as a measure of polarization would conclude that polarization underwent a small decrease following each election. A look at the measure $P\left(F,x^{*}\right)$ for different $x^{*}$ reveals, however, that this overall decrease masks a more complicated process: polarization around left-wing positions decreased while polarization around right-wing positions increased. 

%between 2004 and 2012, polarization around various left-of-center positions was mostly unchanged, while polarization around right wing positions was growing. To put it differently, during this period right-of-center voters were becoming progressively more distant ideologically from the rest of the electorate, while the distance between left-of-center voters and other voters was fairly stable. From 2012 onwards, on the other hand, polarization was growing around all positions -- informally, voters of all ideological positions were becoming more distant from other voters. 

Third, the model also enables the researcher to do the opposite exercise: to find a cleavage point $x^{*}$ around which polarization has increased the most between two given years. For example, data shows that polarization around all positions was increasing after 2012. But what is the main driver of the increased polarization, that is, where does the most important divide lie? The paper shows that the largest increases of polarization (in percentage terms) occurred around positions slightly to the left of the center of the scale.

%between 2008 and 2012, the variance of voters' positions on the left-right spectrum was almost unchanged. This might seem to suggest that there was no change in how polarized the electorate was. However, $P\left(F,x^{*}\right)$ reveals a more nuanced picture: between 2008 and 2012, polarization around every position on the left half of the scale has decreased, while polarization around every position to the right of the middle has increased. Hence, by being flexible about the location of the central point, this measure of polarization allows a researcher to uncover dynamics that standard measure of polarization do not reveal.

Furthermore, I show how this definition is related to several features of polarization. First, it helps link ideological polarization described above to affective polarization, that is, increased dislike towards members of the opposing political group. Past research \citep{rogowski2016ideology,bougher2017correlates,webster2017ideological,orr2020policy} suggests that increased ideological divide may drive affective polarization. This paper provides a formal model consistent with these empirical results. I model a setting in which $x^{*}$ serves as a boundary between two political groups. Each voter's animosity towards the opposite group is an increasing function of ideological distance between her and the average member of that group. The paper shows that increased ideological polarization around $x^{*}$ (according to the partial order described above) implies increased average level of animosity.

At the same time, there is some evidence \citep{han2023issue} that increased salience of divisive issues may be related to increase in affective polarization. The framework for analyzing polarization developed here provides a mechanism consistent with this observation. I consider a formal model in which a voter's political position on the left-right spectrum is a weighted average of her positions on two issues. The paper shows that an increase in the weight of the more divisive issue implies increased ideological polarization as defined in this paper, and hence also increased affective polarization.

\section*{A Model of Polarization }

Consider a set of voters, and a set $X\subset\mathbb{R}$ of policies. Depending on the application, $X$ may be continuous or discrete. Let $F$ be the cumulative distribution of voters' positions over $X$. To fix ideas, we will say that the higher the value of $x$, the more right-wing the voter is. 

Let $x^{*}\in X$ be the policy that a researcher considers centrist. For the rest of the analysis, we will assume that $x^{*}$ is in the interior of $X$. One might think of centrist voters as those whose positions are close to $x^{*}$, that is, fall into some interval $\left[\underline{x},\overline{x}\right]$ that includes $x^{*}$. Polarization then involves a reduction in the fraction of voters whose policies belong to the centrist interval $\left[\underline{x},\overline{x}\right]$, and an increase in the fraction of voters whose policies lie outside of it. However, such a definition relies on an arbitrary choice of the boundaries $\underline{x}$ and $\overline{x}$. Instead, we can define polarization as a reduction in the fraction of voters who are centrists, \emph{whichever way the boundaries of the center are set:}
\begin{defn}
\label{def:polarisation}Consider two distributions $F$ and $\hat{F}$, and a policy $x^{*}\in X$. We say that $\hat{F}$ dominates $F$ in polarization around \emph{$x^{*}$ }if and only if for any interval $\left[\underline{x},\overline{x}\right]$ that includes $x^{*}$, the share of voters whose positions belong to $\left[\underline{x},\overline{x}\right]$ is weakly smaller under $\hat{F}$ than under $F$.
\end{defn}
Thus, increased polarization means a reduction in the fraction of centrists, and an increase in the fraction of radicals, wherever the dividing lines between the former and the latter are set. A downside of this definition is that it is difficult to work with -- by looking at a pair of distributions, it is not easy to say whether one dominates another in polarization around some $x^{*}$. However, the following result (the proof of which, like those of other propositions, is in the Appendix) provides a simple necessary and sufficient condition to establish this:
\begin{prop}
\label{prop: centre} $\hat{F}$ dominates $F$ in polarization around \emph{$x^{*}$} if and only if both of the following statements are true: (i) $\hat{F}\left(x\right)-F\left(x\right)\geq0$ for all $x<x^{*}$, and (ii) $\hat{F}\left(x\right)-F\left(x\right)\leq0$ for all $x\geq x^{*}$.
\end{prop}
Hence, $\hat{F}$ implies higher polarization around $x^{*}$ than $F$ if $\hat{F}\left(x\right)-F\left(x\right)$ crosses zero exactly once, at $x^{*}$. Equivalently, polarization around $x^{*}$ is higher under $\hat{F}$ than under $F$ if $F$ first-order stochastically dominates $\hat{F}$ for $x<x^{*}$, while the reverse holds for $x>x^{*}$. Similar single-crossing conditions on cumulative distribution functions have been used in economics literature \citep{diamond1974increases,hammond1974simplifying,johnson2006simple,drugov2020noise}, though not in the context of political polarization. If $F$ and $\hat{F}$ have the same mean, dominance in polarization implies a mean-preserving spread, but in general the distributions do not have to have the same mean.

A few things are worth noting about Definition \ref{def:polarisation} and Proposition \ref{prop: centre}. First, the concept of polarization that they introduce is specific to a given central position: it describes polarization around a particular point $x^{*}$. For example, polarization may increase around the median position. Or instead there may be an increase in polarization around a moderate right position, that is, between the extreme right and other voters. As will be shown later, this flexibility is helpful for determining the point, if any, around which polarization occurs.

Second, they only introduce a partial order on the set of distributions: not every pair of distributions can be compared in terms of polarization this way. Specifically, if the difference between two distributions crosses zero multiple times, neither distribution dominates another in polarization around any $x^{*}$. It would, however, be useful to have a numerical measure of polarization that captures the spirit of the disappearance of the center while allowing for comparisons between any two distributions. An intuitive way to define it is to say that polarization is higher under $\hat{F}$ than under $F$ whenever the condition $\hat{F}\left(x\right)-F\left(x\right)\geq0$ ``tends to be hold'' for $x\leq x^{*}$, and ``tends to fail'' for $x\geq x^{*}$. Formally, we can define the degree of polarization as follows:
\begin{defn}
\label{def: polarisation_numerical}The degree of polarization of distribution $F$ around $x^{*}$ is given by
\[
P\left(F,x^{*}\right):=\frac{\int_{x<x^{*}}F\left(x\right)dx}{x^{*}-\min\left\{ X\right\} }-\frac{\int_{x\geq x^{*}}F\left(x\right)dx}{\max\left\{ X\right\} -x^{*}}+1.
\]
Under this definition, the degree of polarization around $x^{*}$ increases whenever $F\left(x\right)$ tends to become higher for $x\leq x^{*}$, and tends to become lower for $x\geq x^{*}$. As defined here, $P\left(F,x^{*}\right)$ has the following useful properties:
\end{defn}
\begin{prop}
\label{prop: numerical_properties}$P\left(F,x^{*}\right)\in\left[0,1\right]$ for any $F$ and any $x^{*}\in X$. Furthermore, for any distributions $F,\hat{F}$ and any $x^{*}\in X$, if $\hat{F}$ dominates $F$ in polarization around \emph{$x^{*}$, then $P\left(\hat{F},x^{*}\right)>P\left(F,x^{*}\right)$.}
\end{prop}
The first statement says that $P\left(F,x^{*}\right)$ has the same scale regardless of $x^{*}$, which makes it easier to compare polarization around different central points. It is a consequence of dividing the integrals by the width of their support. The second says that if one distribution dominates another in polarization around some central point, then it has a higher degree of polarization around it. Hence, the numerical measure $P\left(F,x^{*}\right)$ is consistent with the partial order introduced by Definition \ref{def:polarisation}.

\section*{Application: Polarization of the US Electorate}

This section illustrates how the proposed measure $P\left(F,x^{*}\right)$ can be used to gain insights about electoral polarization. The analysis draws on data from the American National Election Studies database. It uses two measures of ideology: voters' position on a left-right political axis, and on a liberal-conservative axis. Table \ref{tab:shares} in the Appendix presents the distribution of respondents' answers when asked to position themselves on a scale from 0 to 10, where 0 means the left and 10 means the right, for several years. Table \ref{tab:shares_con_lib} in the Appendix presents respondents' self-placement on a scale from 1 to 7, where 1 means extremely liberal, and 7 means extremely conservative.

\subsection*{Polarization over time}

Consider first the question of how polarization has evolved over time, starting from the left-right ideological spectrum. Table \ref{tab:shares} in the Appendix shows that the variance of US voters' ideological positions remained relatively stable between 2004 and 2012, and has been increasing since then. However, variance and similar distributional moments, by construction, only measure polarization around the average position. 

\begin{figure}
\centering{}%
\begin{minipage}[t]{0.47\columnwidth}%
\begin{center}
\includegraphics[scale=0.5]{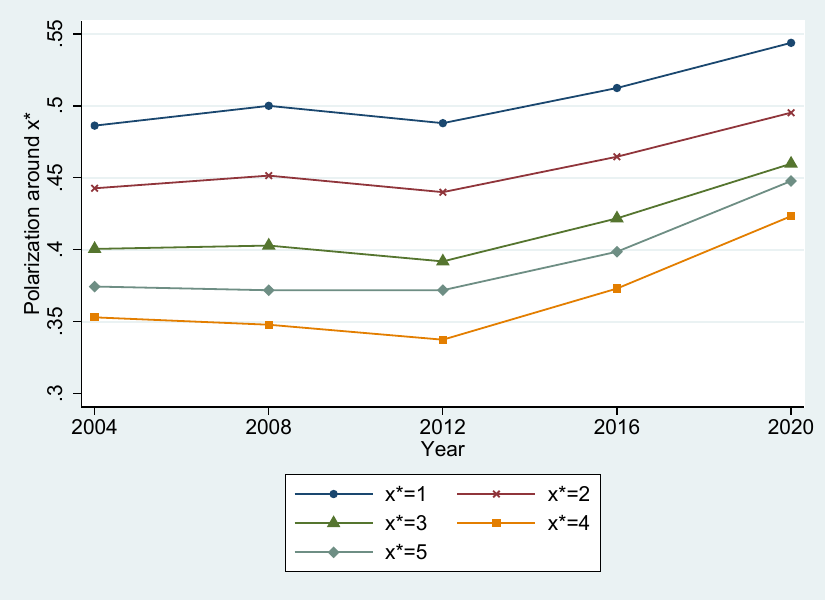}
\par\end{center}%
\end{minipage}%
\begin{minipage}[t]{0.47\columnwidth}%
\begin{center}
\includegraphics[scale=0.5]{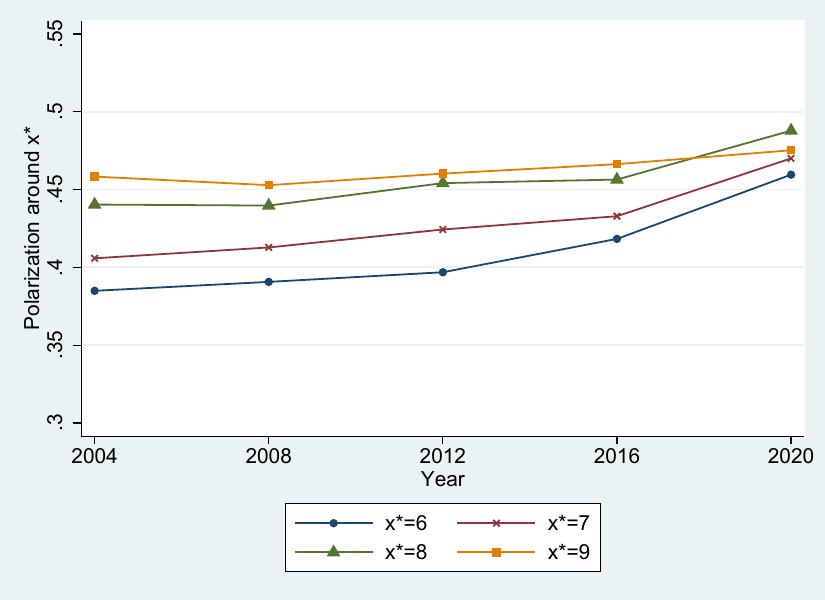}
\par\end{center}%
\end{minipage}\caption{\label{fig:Evolution-of-polarization}Evolution of polarization of US electorate between 2004 and 2020 on the left-right axis. Left: polarization around positions $x^{*}\protect\leq5$. Right: polarization around positions $x^{*}>5$.}
\end{figure}

On the other hand, the degree of polarization $P\left(F,x^{*}\right)$ introduced in this paper allows analysis of polarization centered at any point $x^{*}$ of interest to a researcher. Figure \ref{fig:Evolution-of-polarization} plots $P\left(F,x^{*}\right)$ for different central points across the years from 2004 to 2020. The left panel shows the degree of polarization around different values of $x^{*}$ between 1 and 5, that is, around left-of-center positions and the middle position. The right panel shows the degree of polarization around $x^{*}\geq6$, that is, around right-of-center positions. The figure shows that the degree of polarization around left-of-center positions decreased somewhat between 2004 and 2012 for all $x^{*}\in\left[2,5\right]$, and increased only slightly for $x^{*}=1$. After 2012, however, the degree of polarization rose steeply for all $x^{*}\leq5$. By contrast, the degree of polarization around right-of-center positions rose gradually across the entire time.\footnote{For $x^{*}\in\left[8,9\right]$, the measure $P\left(F,x^{*}\right)$ decreased slightly between 2004 and 2008, but overall between 2004 and 2012 it became larger.} 

%One can see from Figure \ref{fig:Evolution-of-polarization} that the increase in the degree of polarization has happened in different ways for left-wing and right-wing central points. As the left panel shows, the degree of polarization around left-of-center positions -- that is, the distance between left-leaning voters and the rest of the electorate -- was not increasing until 2012. In fact, $P\left(F,x^{*}\right)$ decreased between 2004 and 2012 for all $x^{*}\in\left[2,5\right]$, and increased only slightly for $x^{*}=1$. After 2012, however, the degree of polarization rose steeply for all $x^{*}\leq5$. On the other hand, the degree of polarization around right-of-center positions, shown on the right, grew gradually and consistently throughout the period.\footnote{For $x^{*}\in\left[8,9\right]$, the measure $P\left(F,x^{*}\right)$ decreased slightly between 2004 and 2008, but overall between 2004 and 2012 it became larger.} 

\begin{figure}
\centering{}%
\begin{minipage}[t]{0.47\columnwidth}%
\begin{center}
\includegraphics[scale=0.5]{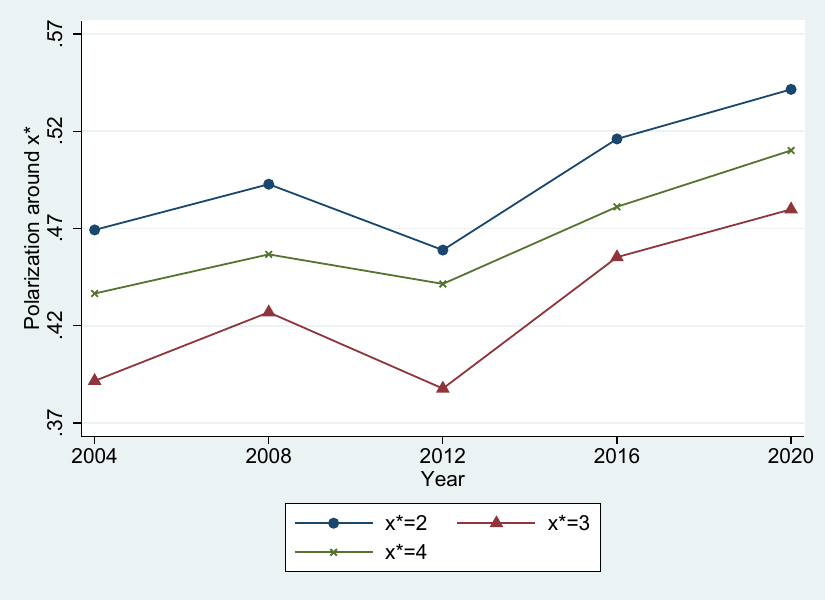}
\par\end{center}%
\end{minipage}%
\begin{minipage}[t]{0.47\columnwidth}%
\begin{center}
\includegraphics[scale=0.5]{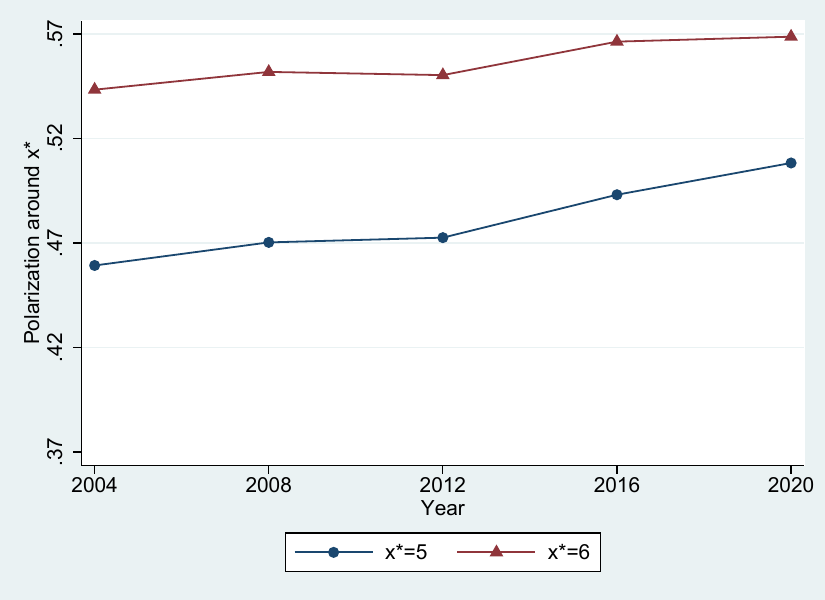}
\par\end{center}%
\end{minipage}\caption{\label{fig:Evolution-of-polarization-lib-con}Evolution of polarization of US electorate between 2004 and 2020 on the liberal-conservative axis. Left: polarization around positions $x^{*}\protect\leq4$. Right: polarization around positions $x^{*}>4$.}
\end{figure}

Turning to the liberal-conservative spectrum, Table \ref{tab:shares_con_lib} shows that the overall variance of positions fluctuated between 2004 and 2012, and increased subsequently. At the same time, the left panel of Figure \ref{fig:Evolution-of-polarization-lib-con} shows that, similarly to the earlier result, the degree of polarization around central points on the liberal side of the spectrum fluctuated somewhat until 2012, and increased sharply thereafter. On the other hand, the degree of polarization around positions on the conservative side of the spectrum, shown on the right, underwent a slow but constant increase.

In sum, while traditional measures of polarization indicate that the US electorate has become more polarized, the $P\left(F,x^{*}\right)$ metric additionally suggests that the trajectory of the increase differs depending on the relevant central point. The degree of polarization around conservative or right-of-center positions increased steadily throughout the period. On the other hand, the sustained increase degree of polarization around liberal or left-of-center positions only began after 2012 but was sharper. By focusing on polarization around particular positions instead of looking at polarization as a single characteristic of the entire distribution, the measure $P\left(F,x^{*}\right)$ reveals these divergent patterns, enabling the researcher to obtain a more complete picture of polarization.

As a particular example, consider the change in voter distribution between 2008 and 2012 on the left--right scale. Variance declined by just 0.52\%, and the share of voters between positions 4 and 6 rose slightly from 48.4\% to 50.4\%. A researcher relying on these conventional indicators would find a modest decline in polarization. However, these measures are, by construction, centered on an arbitrarily selected point: respectively, the mean, and the position 5. If one examines polarization around other positions, Figure \ref{fig:Evolution-of-polarization} shows that between 2008 and 2012, $P\left(F,x^{*}\right)$ has decreased for all $x^{*}<5$, increased for all $x^{*}>5$, and remained unchanged for $x^{*}=5$. Hence, the decrease in variance masks an increase in polarization around right-of-center positions, counteracted by a decrease in polarization around left-of-center positions.

\subsection*{Polarization during elections}

Researchers may also wish to examine shorter-term changes in polarization, such as those occurring during election years. Table \ref{tab:shares_con_lib_pre_post} in the Appendix shows the distribution of voters' positions on the liberal-conservative spectrum before and after the 2004 and 2016 elections. In both years, variance of positions decreased. Does this imply a reduction in polarization?

\begin{figure}
\centering{}%
\begin{minipage}[t]{0.47\columnwidth}%
\begin{center}
\includegraphics[scale=0.5]{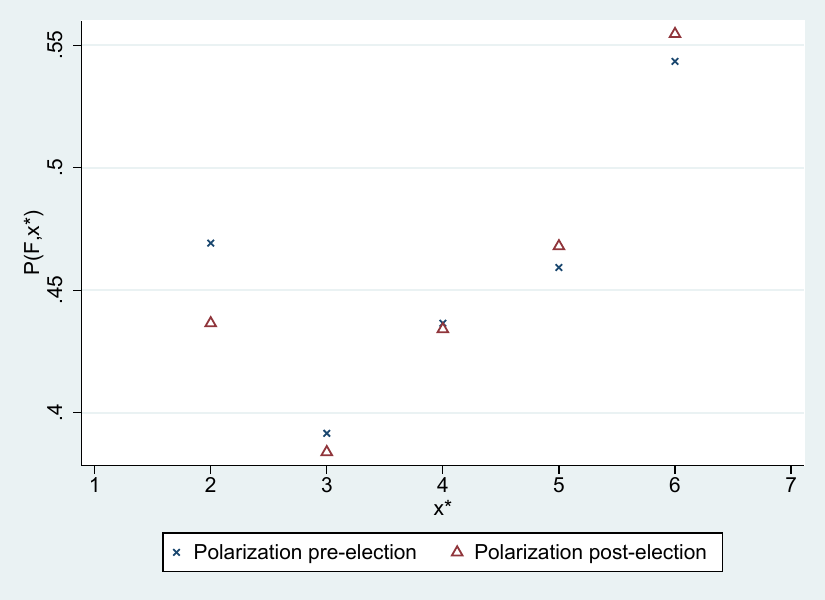}
\par\end{center}%
\end{minipage}%
\begin{minipage}[t]{0.47\columnwidth}%
\begin{center}
\includegraphics[scale=0.5]{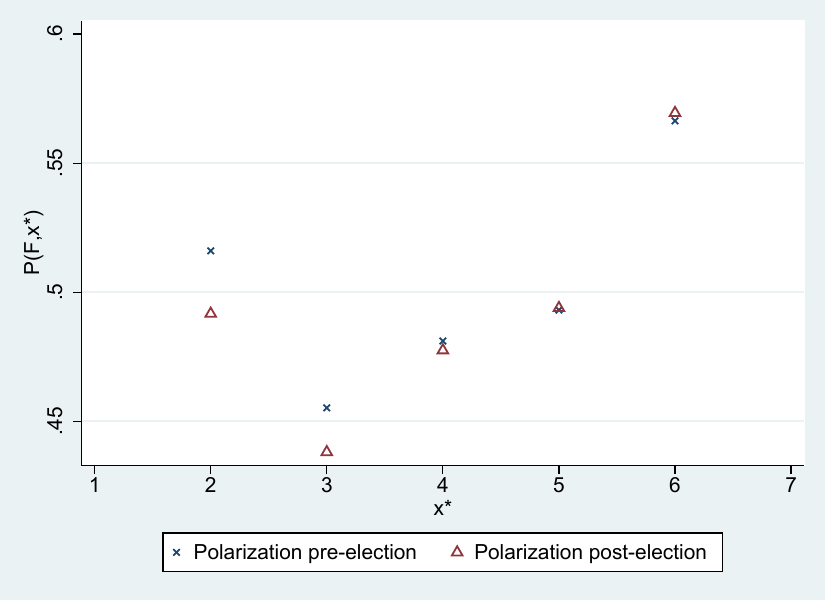}
\par\end{center}%
\end{minipage}\caption{\label{fig:before-after election}Polarization around different positions before and after the election. Left: 2004 election. Right: 2016 election. Note the different scales.}
\end{figure}

Figure \ref{fig:before-after election} addresses this question by plotting $P\left(F,x^{*}\right)$ before and after the 2004 (left panel) and the 2016 (right panel) elections. In both cases, $P\left(F,x^{*}\right)$ decreased after the election for $x^{*}\leq4$, and increased for $x^{*}>4$. Thus, the degree of polarization between liberal voters and all other voters has decreased, while the degree of polarization between conservative voters and other voters has increased.

%To put it differently, voters on the liberal side of the spectrum have become less different from other voters, while voters on the conservative side have become more different. 

\subsection*{Finding cleavages}

The preceding analysis shows how a researcher that starts the analysis with some position of interest $x^{*}$ can measure the degree of polarization around it. Suppose, however, that the researcher wants to take the opposite path: instead of starting with a given central point, she wants to determine the positions around which polarization principally occurs.

\begin{figure}
\centering{}%
\begin{minipage}[t]{0.47\columnwidth}%
\begin{center}
\includegraphics[scale=0.5]{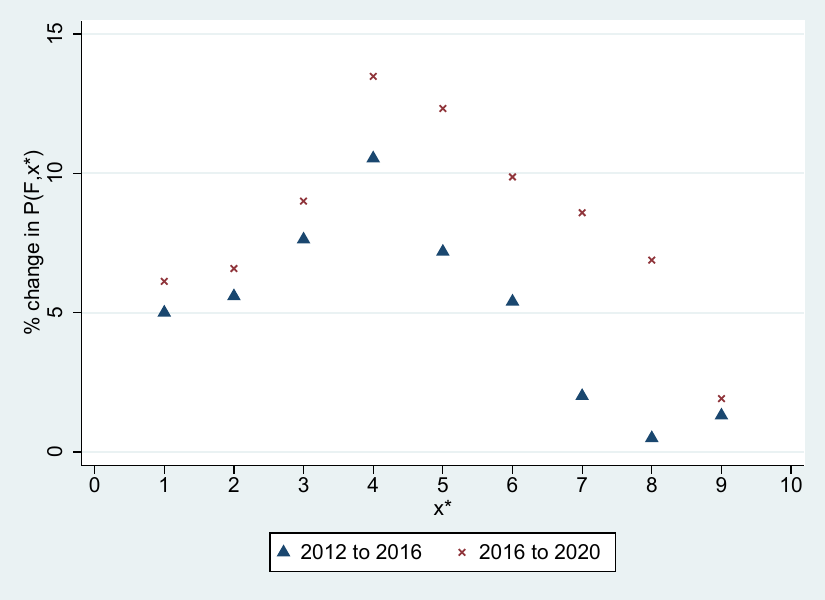}
\par\end{center}%
\end{minipage}%
\begin{minipage}[t]{0.47\columnwidth}%
\begin{center}
\includegraphics[scale=0.5]{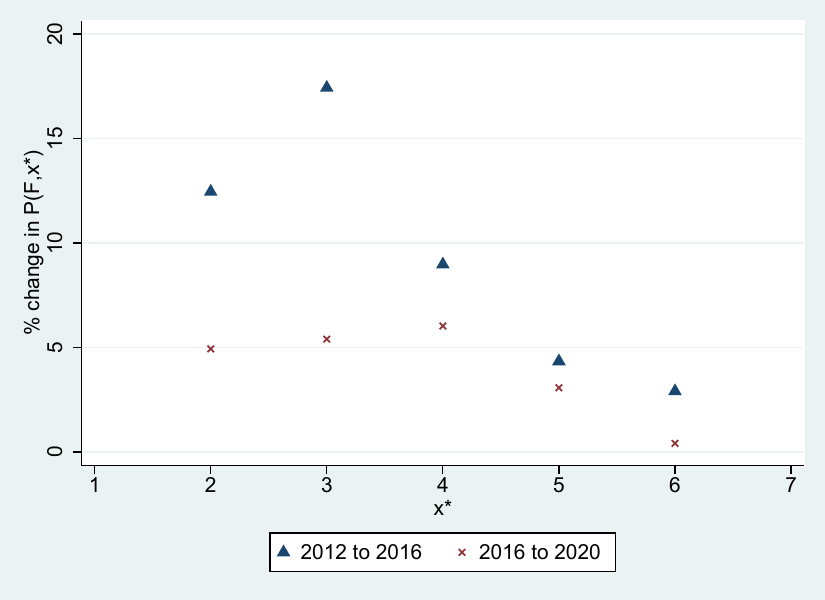}
\par\end{center}%
\end{minipage}\caption{\label{fig:Cleavages post2012}Percentage changes in polarization on the left-right axis (left) and on the liberal-conservative axis (right) between 2012 and 2020 for different central positions $x^{*}$. Note the different scales.}
\end{figure}

Consider the period after 2012, during which polarization increased across the spectrum. Where was this increase most pronounced? Using the left--right scale, the left panel of Figure \ref{fig:Cleavages post2012} plots the percentage changes in $P\left(F,x^{*}\right)$ from 2012 to 2016 and from 2016 to 2020. The largest increases occurred around $x^{*}=4$, that is, slightly to the left of the middle of the scale. We can therefore conclude that the relevant cleavage during this period was between left-wing voters and the rest of the electorate. Similarly, the right panel of Figure \ref{fig:Cleavages post2012} plots the percentage changes in $P\left(F,x^{*}\right)$ on the liberal-conservative spectrum. Again, the largest increases occurred around positions in the middle of the scale or slightly on the liberal side of it. 

\part*{Discussion\label{sec:Discussion}}

\paragraph{Ideological polarization and affective polarization.}

A growing body of research focuses on affective polarization: the dislike towards members of the opposite political group.\footnote{See \citet{iyengar2019origins} for an overview.} How does ideological polarization developed here relate to affective polarization?

Suppose the ideological spectrum is divided at $x^{*}$ into two groups, the left ($L$) and the right ($R$). Voters with $x<x^{*}$ belong to $L$, and voters with $x>x^{*}$ belong to $R$. Suppose that a voter forms a more negative opinion about the opposing group if that group is on average more dissimilar to her. Formally, suppose a voter with position $x\in X$ belonging to group $i\in\left\{ L,R\right\} $ experiences animosity towards the opposing group $j\neq i$ given by $g\left(\left|x-m_{j}\right|\right)$, where $g\left(\cdot\right)$ is some increasing function, $\left|\cdot\right|$ is the absolute value, and $m_{j}$ is the position of the average member of group $j$. Let the overall level of affective polarization be the average value of animosity across all voters. Then we can show the following result:
\begin{prop}
\label{prop: affective}Take any increasing function $g\left(\cdot\right)$. If $\hat{F}$ dominates $F$ in polarization around \emph{$x^{*}$}, then affective polarization is higher under $\hat{F}$ than under $F$.
\end{prop}
In words, increased ideological polarization  around the boundary between the two political groups implies a higher level of affective polarization between these groups. Note that the reverse does not necessarily hold: affective polarization may also increase through a shift in $F$ that is different from the one described in Definition \ref{def:polarisation}. 

\paragraph{Polarization and salience of divisive issues.}

Some issues are more divisive than others. For example, there is evidence that the US electorate is more divided on social issues than on economic issues \citep{saad2021americans}. Suppose that the position $x\in X$ of a voter is determined by her stance on two issues, one of which is ``divisive'', while the other is ``common-value''. Specifically, suppose that for each voter, we have 
\[
x=\left(1-\alpha\right)c+\alpha d,
\]
where $c$ and $d$ are real numbers, and $\alpha\in\left[0,1\right]$. The numbers $c$ and $d$ represent the voter\textquoteright s position on the common-value and divisive issues, respectively, drawn independently from distributions $G_{c}$ and $G_{d}$. The two issues differ in the following way: $d$ can take any value on $X$, while $c$ can only take values on some interval $\left[\underline{c},\overline{c}\right]\subset X$. In words, the range of voters' positions on the common-value issue is more narrow than the range of positions on the divisive issue. The coefficient $\alpha$ captures the salience of the divisive issue.\footnote{In the limiting case when $\underline{c}$ and $\overline{c}$ are arbitrarily close to each other and equal the expected value of $d$, a decrease in $\alpha$ is equivalent to a ``squeeze'' of the distribution of $x$ in the language of \citet{duclos2004polarization}.} Let $F_{\alpha}$ be the distribution of $x$ for a given $\alpha$. We can show the following result:
\begin{prop}
\label{prop: salience}If $\underline{c}$ and $\overline{c}$ are sufficiently close, then for any $\alpha,\alpha^{\prime}$ such that $\alpha^{\prime}>\alpha$, the distribution $F_{\alpha^{\prime}}$ dominates the distribution $F_{\alpha}$ in polarization around some $x^{*}\in\left[\underline{c},\overline{c}\right]$.
\end{prop}
That is, when voters' views on the common-value issue are sufficiently similar, increased salience of the divisive issue makes the electorate more polarized. Recall that by Proposition \ref{prop: affective}, increased ideological polarization implies increased affective polarization. Hence, greater salience of divisive issues also increases affective polarization.

\section*{Conclusions}

Ideological polarization remains a central topic in political economy, yet its measurement remains contentious. This paper introduces a flexible framework for evaluating polarization around any point of interest. Unlike traditional measures that aggregate polarization into a single summary statistic, this approach enables more granular and detailed analysis. At the same time, it provides a framework for linking ideological polarization, affective polarization, and salience of particular policy issues.

%Ideological polarization is a topic of considerable research, which, however, is complicated by a lack of consensus on how to define polarization. This paper has proposed a way of comparing distributions of voters' ideological positions in terms of polarization around a given point. Because this measure is flexible in determining the point around which polarization occurs, it allows a richer and more nuanced analysis of polarization, compared to traditional measures. At the same time, it provides a framework for linking ideological polarization, affective polarization, and salience of particular policy issues.

\begin{singlespace}
\bibliography{polarisation}

\end{singlespace}

\newpage
\appendix
\setcounter{page}{1}
%\counterwithin{table}{section}
\setcounter{table}{0} 
\renewcommand{\thetable}{A\arabic{table}} 

\section*{Appendix}

\subsection*{Additional Tables}

\begin{table}[H]
\centering{}%
\begin{tabular}{|c|ccccc|}
\hline 
\multirow{2}{*}{\textbf{Political position}} & \multicolumn{5}{c|}{\textbf{Share in...}}\tabularnewline
 & \textbf{2004} & \textbf{2008} & \textbf{2012} & \textbf{2016} & \textbf{2020}\tabularnewline
\hline 
\textbf{0 (Left)} & 1.169\% & 1.641\% & 1.790\% & 2.734\% & 4.469\%\tabularnewline
\textbf{1} & 2.682\% & 2.478\% & 1.769\% & 2.623\% & 2.784\%\tabularnewline
\textbf{2} & 4.478\% & 3.880\% & 4.359\% & 4.845\% & 5.825\%\tabularnewline
\textbf{3} & 6.211\% & 5.126\% & 5.405\% & 6.007\% & 7.018\%\tabularnewline
\textbf{4} & 7.316\% & 6.516\% & 6.802\% & 6.717\% & 7.728\%\tabularnewline
\textbf{5} & 28.280\% & 30.580\% & 33.277\% & 29.168\% & 25.292\%\tabularnewline
\textbf{6} & 9.868\% & 11.313\% & 10.281\% & 10.594\% & 7.998\%\tabularnewline
\textbf{7} & 14.086\% & 12.148\% & 11.607\% & 9.796\% & 10.074\%\tabularnewline
\textbf{8} & 12.395\% & 10.739\% & 9.689\% & 11.612\% & 10.935\%\tabularnewline
\textbf{9} & 5.558\% & 6.731\% & 5.207\% & 6.815\% & 5.366\%\tabularnewline
\textbf{10 (Right)} & 7.956\% & 8.849\% & 9.815\% & 9.090\% & 12.511\%\tabularnewline
\hline 
\textbf{Average position} & 5.875 & 5.925 & 5.858 & 5.803 & 5.723\tabularnewline
\textbf{Variance of positions} & 5.336 & 5.426 & 5.397 & 6.108 & 7.376\tabularnewline
\hline 
\end{tabular}\caption{\label{tab:shares}Shares of respondents by self-reported position on a 0-10 left-right axis, by year. Source: ANES dataset, electionstudies.org/data-tools/anes-continuity-guide/\#left-right. Answers are weighted using the weights provided by ANES. Respondents who did not know how to answer or refused to answer are omitted.}
\end{table}

\begin{table}[H]
\centering{}%
\begin{tabular}{|c|ccccc|}
\hline 
\multirow{2}{*}{\textbf{Political position}} & \multicolumn{5}{c|}{\textbf{Share in...}}\tabularnewline
 & \textbf{2004 } & \textbf{2008} & \textbf{2012} & \textbf{2016} & \textbf{2020}\tabularnewline
\hline 
\textbf{1 (Extremely liberal)} & 3.04\% & 3.68\% & 3.14\% & 4.27\% & 5.04\%\tabularnewline
\textbf{2} & 11.63\% & 13.03\% & 11.16\% & 15.00\% & 15.76\%\tabularnewline
\textbf{3} & 10.39\% & 11.80\% & 11.68\% & 12.29\% & 12.10\%\tabularnewline
\textbf{4} & 33.49\% & 29.06\% & 34.42\% & 27.11\% & 27.03\%\tabularnewline
\textbf{5} & 15.82\% & 15.96\% & 15.59\% & 15.44\% & 12.04\%\tabularnewline
\textbf{6} & 21.62\% & 22.40\% & 19.40\% & 21.20\% & 21.70\%\tabularnewline
\textbf{7 (Extremely conservative)} & 4.01\% & 4.07\% & 4.62\% & 4.69\% & 6.32\%\tabularnewline
\hline 
\textbf{Average position} & 4.283 & 4.241 & 4.248 & 4.168 & 4.156\tabularnewline
\textbf{Variance of positions} & 2.147 & 2.334 & 2.132 & 2.503 & 2.738\tabularnewline
\hline 
\end{tabular}\caption{\label{tab:shares_con_lib}Shares of respondents by self-reported position on a 1-7 liberal-conservative scale, by year, pre-election. Source: ANES dataset, https://electionstudies.org/data-tools/anes-continuity-guide/\#liberal-conservative. Answers are weighted using the weights provided by ANES. Respondents who did not know how to answer or refused to answer are omitted.}
\end{table}

\begin{table}[H]
\centering{}%
\begin{tabular}{|c|c|c|c|c|}
\hline 
\multirow{2}{*}{\textbf{Political position}} & \multicolumn{2}{c|}{\textbf{Share in 2004}} & \multicolumn{2}{c|}{\textbf{Share in 2016}}\tabularnewline
\cline{2-5}
 & \textbf{Before} & \textbf{After} & \textbf{Before} & \textbf{After}\tabularnewline
\hline 
\textbf{1 (Extremely liberal)} & 3.04\% & 2.54\% & 4.27\% & 3.62\%\tabularnewline
\textbf{2} & 11.63\% & 10.84\% & 15.00\% & 14.52\%\tabularnewline
\textbf{3} & 10.39\% & 14.31\% & 12.29\% & 13.42\%\tabularnewline
\textbf{4} & 33.49\% & 32.44\% & 27.11\% & 29.11\%\tabularnewline
\textbf{5} & 15.82\% & 16.68\% & 15.44\% & 14.35\%\tabularnewline
\textbf{6} & 21.62\% & 19.88\% & 21.20\% & 20.70\%\tabularnewline
\textbf{7 (Extremely conservative)} & 4.01\% & 3.30\% & 4.69\% & 4.28\%\tabularnewline
\hline 
\textbf{Average position} & 4.283 & 4.227 & 4.168 & 4.153\tabularnewline
\textbf{Variance of positions} & 2.147 & 2.013 & 2.503 & 2.375\tabularnewline
\hline 
\end{tabular}\caption{\label{tab:shares_con_lib_pre_post}Shares of respondents by self-reported position on a 1-7 liberal-conservative scale, by year. Before and After refer to surveys taken before and after the election, respectively. Source: ANES dataset, https://electionstudies.org/data-tools/anes-continuity-guide/\#liberal-conservative. Answers are weighted using the weights provided by ANES. Respondents who did not know how to answer or refused to answer are omitted.}
\end{table}

\subsection*{Proofs}

\paragraph{Proof of Proposition \ref{prop: centre}.}

For any interval $\left[\underline{x},\overline{x}\right]$, the fraction of voters whose positions belong to that interval equals $F\left(\overline{x}\right)-F\left(\underline{x}\right)$. Hence, polarization around $x^{*}$ is higher under $\hat{F}$ than under $F$ if and only if
\begin{equation}
\hat{F}\left(\overline{x}\right)-\hat{F}\left(\underline{x}\right)\leq F\left(\overline{x}\right)-F\left(\underline{x}\right)\text{ for all }\underline{x},\overline{x}\text{ such that }\underline{x}\leq x^{*}\leq\overline{x}\label{eq:centre reduction}
\end{equation}
To prove the proposition, we need to prove that for any $x^{*}\in X$, the two conditions given there hold if and only if (\ref{eq:centre reduction}) holds.

Take some $x^{*}\in X$. To prove one direction of the statement, suppose (\ref{eq:centre reduction}) holds. When $\underline{x}=\min\left\{ X\right\} $, we have $\hat{F}\left(\underline{x}\right)=F\left(\underline{x}\right)=0$, which together with (\ref{eq:centre reduction}) implies that $\hat{F}\left(\overline{x}\right)\leq F\left(\overline{x}\right)$ for every $\overline{x}\geq x^{*}$. When $\overline{x}=\max\left\{ X\right\} $, we have $\hat{F}\left(\overline{x}\right)=F\left(\overline{x}\right)=1$, which together with (\ref{eq:centre reduction}) implies that $\hat{F}\left(\underline{x}\right)\geq F\left(\underline{x}\right)$ for every $\underline{x}\leq x^{*}$.

To prove the other direction, suppose that $\hat{F}\left(\underline{x}\right)\geq F\left(\underline{x}\right)$ for all $\underline{x}\leq x^{*}$, and $\hat{F}\left(\overline{x}\right)\leq F\left(\overline{x}\right)$ for all $\overline{x}\geq x^{*}$. Subtracting the first inequality from the second, we obtain (\ref{eq:centre reduction}).\qed

\paragraph{Proof of Proposition \ref{prop: numerical_properties}.}

The second statement follows immediately from Proposition \ref{prop: centre}. To prove the first statement, note that the second statement implies that polarization around $x^{*}$ is highest when some fraction of voters $\alpha$ have positions at $\min\left\{ X\right\} $, and the remaining fraction $1-\alpha$ have positions at $\max\left\{ X\right\} $. In that case, $F\left(x\right)=\alpha$ for all $x\in\left(\min\left\{ X\right\} ,\max\left\{ X\right\} \right)$. Hence, for any $x^{*}$ in the interior of $X$, we have $P\left(F,x^{*}\right)=\frac{\alpha\left(x^{*}-\min\left\{ X\right\} \right)}{x^{*}-\min\left\{ X\right\} }-\frac{\alpha\left(\max\left\{ X\right\} -x^{*}\right)}{\max\left\{ X\right\} -x^{*}}+1=1$. At the same time, polarization around $x^{*}$ is lowest when all voters have positions at $x^{*}$. In that case, $F\left(x\right)=0$ for all $x<x^{*}$, and $F\left(x\right)=1$ for all $x\geq x^{*}$. Hence, $P\left(F,x^{*}\right)=\frac{0}{x^{*}-\min\left\{ X\right\} }-\frac{\max\left\{ X\right\} -x^{*}}{\max\left\{ X\right\} -x^{*}}+1=0$.\qed

\paragraph{Proof of Proposition \ref{prop: affective}.}

Suppose polarization around $x^{*}$ is higher under $F$ than under $\hat{F}$ according to Definition \ref{def:polarisation}. Proposition \ref{prop: centre} implies that $F$ first-order stochastically dominates $\hat{F}$ for $x<x^{*}$, and $\hat{F}$ first-order stochastically dominates $F$ for $x>x^{*}$. For each group $j\in\left\{ L,R\right\} $, let $m_{j}$ and $\hat{m}_{j}$ be the average positions of members of the group under distributions $F$ and $\hat{F}$, respectively. By a well-known property of stochastic dominance, we have $m_{L}\geq\hat{m}_{L}$ and $\hat{m}_{R}\geq m_{R}$. 

Let $A\left(F\right)$ and $A\left(\hat{F}\right)$ be the levels of affective polarization under $F$ and under $\hat{F}$, respectively. We have

\begin{align*}
A\left(F\right)= & \int_{x<x^{*}}g\left(m_{R}-x\right)dF\left(x\right)+\int_{x>x^{*}}g\left(x-m_{L}\right)dF\left(x\right)\\
\leq & \int_{x<x^{*}}g\left(m_{R}-x\right)d\hat{F}\left(x\right)+\int_{x>x^{*}}g\left(x-m_{L}\right)d\hat{F}\left(x\right)\\
\leq & \int_{x<x^{*}}g\left(\hat{m}_{R}-x\right)d\hat{F}\left(x\right)+\int_{x>x^{*}}g\left(x-\hat{m}_{L}\right)d\hat{F}\left(x\right)=A\left(\hat{F}\right).
\end{align*}
The first inequality holds due to the stochastic dominance relation between $\hat{F}$ and $F$, and the fact that $g\left(m_{R}-x\right)$ is decreasing in $x$ while $g\left(x-m_{L}\right)$ is increasing in $x$. The second inequality follows from the fact that $m_{L}\geq\hat{m}_{L}$ and $\hat{m}_{R}\geq m_{R}$. Thus, $A\left(\hat{F}\right)\geq A\left(F\right)$.\qed

\paragraph{Proof of Proposition \ref{prop: salience}.}

Since for each voter, $x=\left(1-\alpha\right)c+\alpha d$, the distribution of $x$ is given by 
\[
F\left(x\right)=\Pr\left[\left(1-\alpha\right)c+\alpha d<x\right]=\Pr\left[d<\frac{x-\left(1-\alpha\right)c}{\alpha}\right]=\int_{\underline{c}}^{\overline{c}}G_{d}\left[\frac{x-\left(1-\alpha\right)c}{\alpha}\right]dG_{c}\left(c\right).
\]
For a given $x\in X$, an increase in $\alpha$ increases $F\left(x\right)$ if and only if $\frac{\partial F\left(x\right)}{\partial\alpha}>0$, that is, if and only if
\begin{equation}
\int_{\underline{c}}^{\overline{c}}g_{d}\left[\frac{x-\left(1-\alpha\right)c}{\alpha}\right]\frac{c-x}{\alpha^{2}}dG_{c}\left(c\right)>0,\label{eq:derivative wrt alpha}
\end{equation}
where $g_{d}$ is the density of $d$.

When $x<\underline{c}$, we have $g_{d}\left[\frac{x-\left(1-\alpha\right)c}{\alpha}\right]\frac{c-x}{\alpha^{2}}dG_{c}\left(c\right)>0$ for all $c\in\left[\underline{c},\overline{c}\right]$, so (\ref{eq:derivative wrt alpha}) holds. When $x>\overline{c}$, we have $g_{d}\left[\frac{x-\left(1-\alpha\right)c}{\alpha}\right]\frac{c-x}{\alpha^{2}}dG_{c}\left(c\right)<0$ for all $c\in\left[\underline{c},\overline{c}\right]$, so (\ref{eq:derivative wrt alpha}) does not hold. Now take $x\in\left[\underline{c},\overline{c}\right]$. For these values of $x$, the left-hand side of (\ref{eq:derivative wrt alpha}) is decreasing in $x$ if and only if its derivative with respect to $x$ is negative, that is, if and only if
\begin{align}
 & \int_{\underline{c}}^{\overline{c}}g_{d}^{\prime}\left[\frac{x-\left(1-\alpha\right)c}{\alpha}\right]\frac{c-x}{\alpha^{3}}dG_{c}\left(c\right)-\int_{\underline{c}}^{\overline{c}}g_{d}\left[\frac{x-\left(1-\alpha\right)c}{\alpha}\right]\frac{1}{\alpha^{2}}dG_{c}\left(c\right)<0\nonumber \\
\iff & E\left(g_{d}^{\prime}\left[\frac{x-\left(1-\alpha\right)c}{\alpha}\right]\frac{c-x}{\alpha}\right)<E\left(g_{d}\left[\frac{x-\left(1-\alpha\right)c}{\alpha}\right]\right),\label{eq:condition for salience}
\end{align}
where $E$ denotes the expectation over $G_{c}$. If $\underline{c}$ and $\overline{c}$ are sufficiently close to each other, then $c-x$ is arbitrarily small for all $x\in\left[\underline{c},\overline{c}\right]$. In this case, the left-hand side of (\ref{eq:condition for salience}) is arbitrarily close to zero, while the right-hand side is strictly positive. Hence, (\ref{eq:condition for salience}) holds. Thus, when $\underline{c}$ and $\overline{c}$ are sufficiently close to each other, the left-hand side of (\ref{eq:derivative wrt alpha}) is decreasing in $x$ for all $x\in\left[\underline{c},\overline{c}\right]$. Together with the fact that (\ref{eq:derivative wrt alpha}) holds for all $x<\underline{c}$ and does not hold for all $x>\overline{c}$, this means that there exists $x^{*}\in\left[\underline{c},\overline{c}\right]$ such that (\ref{eq:derivative wrt alpha}) holds if and only if $x<x^{*}$. Hence, an increase in $\alpha$ increases $F\left(x\right)$ for $x<x^{*}$, and decreases $F\left(x\right)$ for $x>x^{*}$, implying an increase in polarization around $x^{*}$.\qed
\end{document}